\documentstyle[12pt]{article}
\begin{document}

\def\ds{\displaystyle}
\def\beq{\begin{equation}}
\def\eeq{\end{equation}}
\def\bea{\begin{eqnarray}}
\def\eea{\end{eqnarray}}
\def\beeq{\begin{eqnarray}}
\def\eeeq{\end{eqnarray}}
\def\ve{\vert}
\def\vel{\left|}
\def\ver{\right|}
\def\nnb{\nonumber}
\def\ga{\left(}
\def\dr{\right)}
\def\aga{\left\{}
\def\adr{\right\}}
\def\lla{\left<}
\def\rra{\right>}
\def\rar{\rightarrow}
\def\nnb{\nonumber}
\def\la{\langle}
\def\ra{\rangle}
\def\ba{\begin{array}}
\def\ea{\end{array}}
\def\tr{\mbox{Tr}}
\def\ssp{{\Sigma^{*+}}}
\def\sso{{\Sigma^{*0}}}
\def\ssm{{\Sigma^{*-}}}
\def\xis0{{\Xi^{*0}}}
\def\xism{{\Xi^{*-}}}
\def\qs{\la \bar s s \ra}
\def\qu{\la \bar u u \ra}
\def\qd{\la \bar d d \ra}
\def\qq{\la \bar q q \ra}
\def\gGgG{\la g^2 G^2 \ra}
\def\q{\gamma_5 \not\!q}
\def\x{\gamma_5 \not\!x}
\def\g5{\gamma_5}
\def\sb{S_Q^{cf}}
\def\sd{S_d^{be}}
\def\su{S_u^{ad}}
\def\rl{\hat{m}_{\ell}}
\def\ss{\hat{s}}
\def\rr{\hat{r}_{K_1}}
\def\sbp{{S}_Q^{'cf}}
\def\sdp{{S}_d^{'be}}
\def\sup{{S}_u^{'ad}}
\def\ssp{{S}_s^{'??}}
\def\sig{\sigma_{\mu \nu} \gamma_5 p^\mu q^\nu}
\def\fo{f_0(\frac{s_0}{M^2})}
\def\ffi{f_1(\frac{s_0}{M^2})}
\def\fii{f_2(\frac{s_0}{M^2})}
\def\O{{\cal O}}
\def\sl{{\Sigma^0 \Lambda}}
\def\es{\!\!\! &=& \!\!\!}
\def\ar{&+& \!\!\!}
\def\ek{&-& \!\!\!}
\def\cp{&\times& \!\!\!}
\def\fhs{Re[FH^*]}
\def\ghs{Re[GH^*]}
\def\bcs{Re[BC^*]}
\def\fgs{Re[FG^*]}
\def\hh{|H|^2}
\def\cc{|C|^2}
\def\gg{|G|^2}
\def\ff{|F|^2}
\def\bb{|B|^2}
\def\aa{|A|^2}
\def\hh{|H|^2}
\def\ee{|E|^2}
\def\rl{\hat{m}_{\ell}}
\def\r{\hat{m}_K}
\def\s{\hat{s}}
\def\ll{\Lambda}

\renewcommand{\textfraction}{0.2}    
\renewcommand{\topfraction}{0.8}

\renewcommand{\bottomfraction}{0.4}
\renewcommand{\floatpagefraction}{0.8}
\newcommand\mysection{\setcounter{equation}{0}\section}
\newcommand{\bra}[1]{\langle {#1}}
\newcommand{\ket}[1]{{#1} \rangle}
\newcommand{\ebar}{{\bar{e}}}
\newcommand{\sbar}{\bar{s}}
\newcommand{\cbar}{\bar{c}}
\newcommand{\bbar}{\bar{b}}
\newcommand{\qbar}{\bar{q}}
\renewcommand{\l}{\ell}
\newcommand{\lbar}{\bar{\ell}}
\newcommand{\psibar}{\bar{\psi}}
\newcommand{\barB}{\overline{B}}
\newcommand{\barK}{\overline{K}}
\newcommand{\thetaK}{\theta_{K_1}}
\newcommand{\onepone}{{1^1P_1}}
\newcommand{\sanpone}{{1^3P_1}}
\newcommand{\kone}{{K_1}}
\newcommand{\barkone}{{\overline{K}_1}}
\renewcommand{\Re}{\mathop{\mbox{Re}}}
\renewcommand{\Im}{\mathop{\mbox{Im}}}
\newcommand{\T}{{\cal T}}
\newcommand{\eff}{{\rm eff}}
\newcommand{\A}{{\cal A}}
\newcommand{\B}{{\cal B}}
\newcommand{\C}{{\cal C}}
\newcommand{\D}{{\cal D}}
\newcommand{\E}{{\cal E}}
\newcommand{\F}{{\cal F}}
\newcommand{\G}{{\cal G}}
\renewcommand{\H}{{\cal H}}
\newcommand{\hats}{\hat{s}}
\newcommand{\hatp}{\hat{p}}
\newcommand{\hatq}{\hat{q}}
\newcommand{\hatm}{\hat{m}}
\newcommand{\hatu}{\hat{u}}
\newcommand{\alphaem}{\alpha_{\rm em}}
\newcommand{\konel}{K_1(1270)}
\newcommand{\koneh}{K_1(1400)}
\newcommand{\barkonel}{\barK_1(1270)}
\newcommand{\barkoneh}{\barK_1(1400)}
\newcommand{\konea}{K_{1A}}
\newcommand{\koneb}{K_{1B}}
\newcommand{\barkonea}{\barK_{1A}}
\newcommand{\barkoneb}{\barK_{1B}}
\newcommand{\mkone}{m_{\kone}}
\newcommand{\konep}{K_1^+}
\newcommand{\konem}{K_1^-}
\newcommand{\konelm}{K_1^-(1270)}
\newcommand{\konehm}{K_1^-(1400)}
\newcommand{\konelp}{K_1^+(1270)}
\newcommand{\konehp}{K_1^+(1400)}
\newcommand{\konelz}{\overline{K}{}^0_1(1270)}
\newcommand{\konehz}{\overline{K}{}^0_1(1400)}
\newcommand{\Bm}{B^-}
\newcommand{\Bz}{\overline{B}{}^0}
\newcommand{\Kstar}{K^*(892)}
\newcommand{\BABAR}{BABAR}
\newcommand{\BELLE}{Belle}
\newcommand{\CLEO}{CLEO}
\newcommand{\leftu}{\gamma^\mu L}
\newcommand{\leftd}{\gamma_\mu L}
\newcommand{\rightu}{\gamma^\mu R}
\newcommand{\rightd}{\gamma_\mu R}
\newcommand{\Br}{{\cal B}}
\newcommand{\sect}[1]{Sec.~\ref{#1}}
\newcommand{\eqref}[1]{(\ref{#1})}
\newcommand{\fig}{FIG.~}
\newcommand{\figs}{FIGs.~}
\newcommand{\tbl}{TABLE~}
\newcommand{\tbls}{TABLEs~}
\newcommand{\errpm}[3]{#1^{+{#2}}_{-{#3}}}
\newcommand{\errpmf}[5]{{#1}^{ +{#2} +{#4} }_{-{#3}-{#5}}}
\newcommand{\lpm}{\l^+\l^-}
\newcommand{\epm}{e^+e^-}
\newcommand{\mupm}{\mu^+\mu^-}
\newcommand{\taupm}{\tau^+\tau^-}
\newcommand{\AFB}{A_{\rm FB}}
\newcommand{\barAFB}{\overline{A}_{\rm FB}}
\newcommand{\GeV}{{\,\mbox{GeV}}}
\newcommand{\MeV}{{\,\mbox{MeV}}}
\newcommand{\degree}{^\circ}
\newcommand{\mB}{m_B}
\newcommand{\SM}{{\rm SM}}
\newcommand{\NP}{{\rm NP}}
\newcommand{\barc}{\bar{c}}
\newcommand{\xipara}{\xi_\parallel^{\kone}}
\newcommand{\xiperp}{\xi_\perp^{\kone}}
\newcommand{\xiparal}{\xi_\parallel^{\konel}}
\newcommand{\xiperpl}{\xi_\perp^{\konel}}
\newcommand{\xiparah}{\xi_\parallel^{\koneh}}
\newcommand{\xiperph}{\xi_\perp^{\koneh}}
\newcommand{\para}{\parallel}
\newcommand{\alphas}{\alpha_s}
\newcommand{\pA}{p_{\kone}}
\newcommand{\lcaption}[2]{\caption{(label:{#2}) #1}\label{#2}}
\newcommand{\Rmunr}{R_{\mu,\rm nr}}
\newcommand{\RdGamma}{R_{d\Gamma/ds,\mu}}
\providecommand{\dfrac}[2]{\frac{\displaystyle
{#1}}{\displaystyle{#2}}}
\def\baeq{\begin{appeq}}     \def\eaeq{\end{appeq}}
\def\baeeq{\begin{appeeq}}   \def\eaeeq{\end{appeeq}}
\newenvironment{appeq}{\beq}{\eeq}
\newenvironment{appeeq}{\beeq}{\eeeq}
%
%
%
%
\title{
         {\Large
                 {\bf
 Forward-backward Asymmetry  and  Branching Ratio of $B \rar K_1 \ell^+ \ell^-$ Transition in
 Supersymmetric Models
                 }
         }
      }

\author{\\
{\small V. Bashiry$^1$\thanks {e-mail: bashiry@ciu.edu.tr},   K.
Azizi$^2$\thanks {e-mail: e146342@metu.edu.tr}\,\,,} \\ {\small $^1$
Engineering Faculty, Cyprus International University,} \\ {\small
Via Mersin 10, Turkey }\\{\small $^2$
Physics Department, Middle East Technical University,}\\
{\small 06531 Ankara, Turkey}}

\date{}

\begin{titlepage}
\maketitle \thispagestyle{empty}

\begin{abstract}
The mass eigen states $K_1(1270)$ and $K_1(1400)$ are mixture of the
strange members of two axial-vector SU(3) octet, $^3P_1(K_1^A)$ and
$^1P_1(K_1^B)$. Taking into account this mixture, the
forward-backward asymmetry and branching ratio of    $B \rar
K_1(1270,1400)\\\ell^+ \ell^-$ transitions are studied in the
framework of different supersymmetric models. It is found that the
results have considerable deviation from the standard model
predictions. Any measurement of these physical observables and their
comparison with the results obtained in this paper can give useful
information about the nature of interactions beyond the standard
model.
\end{abstract}

\end{titlepage}

\section{Introduction}
The Standard Model (SM) explains all experimental predictions well. Despite all the success of SM,
 we can not accept that it is the ultimate theory of nature since there are many questions to be discussed.
   Some issues such as gauge and fermion mass hierarchy, matter- antimatter
 asymmetry, number of generations, the nature
of the dark matter and the unification of fundamental forces can not
be addressed by the SM. In other words, the SM can be considered as
an effective theory of some fundamental theory at low energy.

 One of the most reasonable  extension of
the SM is the Supersymmetry (SUSY) \cite{Ellis}. It is  an important
element in the string theory, which is the most-favored candidate
for unifying the all known interactions including gravity. The SUSY
is assumed to  contribute to overcome the mass hierarchy  problem
between $m_W$ and the Planck scale via canceling the quadratic
divergences in the radiative corrections to the
  Higgs boson mass-squared  \cite{Witten}.

 To verify the SUSY theories, we need to explore the supersymmetric
particles (sparticles). Two types of studies can be conducted to
examine  these sparticles.  In the direct search, the center of mass
energy of colliding particles should be increased to produce SUSY
particles at the TeV scale, hence, it will  be accessible to the
LHC. On the other hand, we can look for SUSY effects, indirectly.
The sparticles can contribute to the transitions at loop level. The
flavor changing neutral current (FCNC) transition of $b\to s$
induced by quantum loop level can be considered as a good condidate
for studying the possible effects of sparticles. For the most recent
studies in this regard see Ref.~\cite{recent} and the references
therein.

The $B \rar K_1 \ell^+ \ell^-$ transition proceeds via the FCNC
transition of $b\to s$ at quark level.  $b\to s$ transition is the
most sensitive and stringiest test for the SM at one loop level,
where, it is forbidden in SM at tree level \cite{R8401,R8402}.
Although, the FCNC transitions have small branching fractions, quite
intriguing results are obtained in ongoing experiments. The
inclusive $B \rar X_s \ell^+ \ell^-$ decay is observed in BaBaR
\cite{R8403} and Belle collaborations. These collaborations have
also announced the measuring exclusive modes $B \rar K \ell^+
\ell^-$ \cite{R8404,R8405,R8406} and $B \rar K^\ast \ell^+ \ell^-$
\cite{R8407}. The obtained  experimental results on these
transitions are in a good consistency with theoretical predictions
\cite{R8408,R8409,R8410} the results of which can be used to
constrain the new physics (NP) effects.

In the present work, calculating the forward-backward asymmetry and
the branching fraction, we investigate the possible effects of
supersymmetric theories on the branching ratio of $B \rar K_1 \ell^+
\ell^-$ transition. Experimentally, the $K_{1} (1270)$ and $K_{1}
(1400)$ are the mixtures of the strange members of the two
axial-vector SU(3) octet $^{3}P_{1}(K_{1}^{A})$ and
$^{1}P_{1}(K_{1}^{B})$. The $K_{1} (1270, 1400)$ and $K_{1}^{A,B}$
states are related to each other as\cite{Hatanaka:2008gu}
\begin{eqnarray}
\pmatrix{|\barkonel \rangle \cr |\barkoneh \rangle} = M \pmatrix{|
\barkonea \rangle \cr | \barkoneb \rangle}, \quad\mbox{with} \quad M
= \pmatrix{ \sin \thetaK & \phantom{-} \cos \thetaK \cr \cos \thetaK
& -\sin \thetaK}. \label{mixing}
\end{eqnarray}
The branching ratio of the $K_{1} (1400)$ case is  smaller than the
$K_{1} (1270)$ \cite{Hatanaka:2008gu}, so  we  consider only $B \rar
K_1(1270) \ell^+\ell^-$. Note that lepton polarization and angular
distribution of this decay in the frame work of SM has recently been
studied in Refs.~\cite{Bashiry:2009wq,Li:2009rc}.

The outline of the paper is
 as follows: In section 2, we calculate the decay amplitude and forward-backward asymmetry
of the $B \rar K_1 \ell^+ \ell^-$ transition within SUSY
models. Section 3 is devoted to the numerical analysis and
discussion of the considered transition as well as our conclusions.
\section{The effective Hamiltonian}\label{sec:Hamiltonian}
The QCD corrected effective Lagrangian for the decays $b\rightarrow
s(d)\ell^{+}\ell^{-}$ can be achieved by integrating out the heavy
quarks and the heavy electroweak bosons in the SUSY model:

\bea \label{e1} {\cal H}_{eff} \es {G_F \alpha V_{tb} V_{ts}^\ast
\over 2\sqrt{2} \pi} \Bigg[ C_9^{eff} (m_b) \bar{s}\gamma_\mu
(1-\gamma_5) b \, \bar{\ell} \gamma^\mu \ell + C_{10} (m_b) \bar{s}
\gamma_\mu (1-\gamma_5) b \, \bar{\ell} \gamma^\mu
\gamma_5 \ell \nnb \\
\ek 2 m_b C_7 (m_b) {1\over q^2} \bar{s} i \sigma_{\mu\nu}
q^{\nu}(1+\gamma_5) b \, \bar{\ell} \gamma^\mu \ell + C_{Q_{1}}
\bar{s} (1+\gamma_5)b ~ \bar{\ell} \ell+ C_{Q_{2}} \bar{s}
(1+\gamma_5)b ~ \bar{\ell}\gamma_5 \ell \Bigg]~,\nnb \\ \eea

where $C_{i}$ are Wilson coefficients and the contributions of SUSY
model are involved via  terms proportional with $C_{Q_{1,2}}$. These
additional terms with respect to the SM   come from the neutral
Higgs bozons(NHBs) exchange diagrams, whose manifest forms and
corresponding Wilson coefficients can be found
in\cite{aslam40,aslam41}. The  $C_{i}$  are calculated in naive
dimensional regularization~(NDR) scheme at the leading order(LO),
next to leading order(NLO) and next-to-next leading order (NNLO) in
the SM\cite{Buras:1994dj}--\cite{NNLL}. $C_9^\eff(\hats) = C_9 +
Y(\hats)$, where $Y(\hats) = Y_{\rm pert}(\hats) + Y_{\rm LD}$
contains both the perturbative part $Y_{\rm pert}(\hats)$ and
long-distance part $Y_{\rm LD}(\hats)$.
$Y(\hats)_{\rm pert}$ is given by \cite{Buras:1994dj}
\begin{eqnarray}
Y_{\rm pert} (\hats) &=& g(\hatm_c,\hats) c_0 \nonumber\\&&
-\frac{1}{2} g(1,\hats) (4 \barc_3 + 4 \barc_4 + 3 \barc_5 +
\barc_6) -\frac{1}{2} g(0,\hats) (\barc_3 + 3 \barc_4) \nonumber\\&&
+\frac{2}{9} (3 \barc_3 + \barc_4 + 3 \barc_5 + \barc_6),
\\
\mbox{with}\quad c_0 &\equiv& \barc_1 + 3\barc_2 + 3 \barc_3 +
\barc_4 + 3 \barc_5 + \barc_6,
\end{eqnarray}
and the function $g(x,y)$ is defined in \cite{Buras:1994dj}.  Here,
$\barc_1$ -- $\barc_6$ are the Wilson coefficients in the leading
logarithmic approximation. The relevant Wilson coefficients are
given  in Refs.~\cite{Ali:1999mm}. $Y(\hats)_{\rm LD}$ involves $B
\to K_1 V(\cbar c)$ resonances \cite{Lim:1988yu}, where $V(\cbar c)$
are the vector charmonium states. We follow
Refs.~\cite{Lim:1988yu,Ali:1991is} and set
\begin{eqnarray}
Y_{\rm LD}(\hats) &=&
 - \frac{3\pi}{\alphaem^2} c_0
\sum_{V = \psi(1s),\cdots} \kappa_V \frac{\hatm_V \Br(V\to
\l^+\l^-)\hat{\Gamma}_{\rm tot}^V}{\hats - \hatm_V^2 + i \hatm_V
\hat{\Gamma}_{\rm tot}^V},
\end{eqnarray}
where $\hat{\Gamma}_{\rm tot}^V \equiv \Gamma_{\rm tot}^V/\mB$ and
$\kappa_V$ takes different value for different exclusive
semileptonic decay. The relevant properties of vector charmonium
states are summarized in Table~\ref{charmonium}.
\begin{table}[tbp]
\caption{Masses, total decay widths and branching fractions of
dilepton decays of vector charmonium states
\cite{Yao:2006px}.}\label{charmonium}
\begin{center}
\begin{tabular}{cclll}
$V$ & Mass[\GeV] &  $\Gamma_{\rm tot}^V$[\MeV]
 &\multicolumn{2}{c}{$\Br(V\to\lpm)$}
\\
\hline $J/\Psi(1S)$ & $3.097$ & $0.093$ & $5.9\times10^{-2}$ & for
$\l=e,\mu$
\\
$\Psi(2S)$   & $3.686$ & $0.327$ & $7.4\times10^{-3}$ & for $\l=e,\mu$ \\
             &         &             & $3.0\times10^{-3}$ & for $\l=\tau$
\\
$\Psi(3770)$ & $3.772$ & $25.2$ & $9.8\times10^{-6}$ & for $\l=e$
\\
$\Psi(4040)$ & $4.040$ & $80$ & $1.1\times10^{-5}$ & for $\l=e$
\\
$\Psi(4160)$ & $4.153$ & $103$ & $8.1\times10^{-6}$ & for $\l=e$
\\
$\Psi(4415)$ & $4.421$ & $62$ & $9.4\times10^{-6}$ & for $\l=e$
\end{tabular}
\end{center}
\end{table}

One has to sandwich the inclusive effective Hamiltonian between
initial hadron state $B(p_B)$ and final hadron state $K_1$ in order
to obtain the matrix element for the exclusive decay $B  \rar K_1
\ell^+ \ell^-$. Following from Eq. (\ref{e1}),  in order to
calculate the decay width and other physical observable of the
exclusive $B \rar K_1 \ell^+ \ell^-$ decay, we need to parameterize
the  matrix elements in terms of formfactors.

The $\barB(p_B)\to \barkone(\pA,\lambda)$ form factors are defined
by\cite{Hatanaka:2008gu}
\begin{eqnarray}
\lefteqn{\bra{\barkone(\pA,\lambda)}|\bar{s} \gamma_\mu (1-\gamma_5)
b|\ket{\barB(p_B)}}
\quad&&\nonumber\\
&=&
 -i \frac{2}{m_B + \mkone} \epsilon_{\mu\nu\rho\sigma}
\varepsilon_{(\lambda)}^{*\nu} p_B^\rho \pA^\sigma A^{\kone}(q^2)
\nonumber
\\
&& -\left[ (m_B + \mkone)\varepsilon_\mu^{(\lambda)*}
V_1^{\kone}(q^2) - (p_B + \pA)_\mu (\varepsilon_{(\lambda)}^* \cdot
p_B)
 \frac{V_2^{\kone}(q^2)}{m_B + \mkone}
\right] \nonumber\\&& +2 \mkone \frac{\varepsilon_{(\lambda)}^*
\cdot p_B}{q^2} q_\mu \left[
 V_3^{\kone}(q^2) - V_0^{\kone}(q^2)
\right], \label{formfactor1}
\\
\lefteqn{\bra{\barkone(\pA,\lambda)}|\bar{s} \sigma_{\mu\nu} q^\nu
(1+\gamma_5) b|\ket{\barB(p_B)}}
\quad&&\nonumber\\
&=& 2T_1^{\kone}(q^2) \epsilon_{\mu\nu\rho\sigma}
\varepsilon_{(\lambda)}^{*\nu} p_B^\rho  \pA^\sigma
\nonumber\\
&& -i T_2^{\kone}(q^2) \left[
 (m_B^2 - \mkone^2) \varepsilon^{(\lambda)}_{*\mu}
-(\varepsilon_{(\lambda)}^{*}\cdot q)
 (p_B + \pA)_\mu
\right] \nonumber\\&& - iT_3^{\kone}(q^2)
(\varepsilon_{(\lambda)}^{*} \cdot q) \left[ q_\mu -
\frac{q^2}{m_B^2 - \mkone^2} (\pA + p_B)_\mu \right],
\label{formfactor2}
\end{eqnarray}
where $q \equiv p_B - \pA=p_{\ell^+}+p_{\ell^-}$. By multiplying
both sides of Eq.~(\ref{formfactor1}) with $q^\mu$, one can obtain
the expression in terms of form factors for
$\lefteqn{\bra{\barkone(\pA,\lambda)}|\bar{s}(1+\gamma_5)
b|\ket{\barB(p_B)}} \quad$.
\begin{eqnarray}
\lefteqn{\bra{\barkone(\pA,\lambda)}|\bar{s}(1+\gamma_5)
b|\ket{\barB(p_B)}}
\quad&&\nonumber\\
&=&
 \frac{1}{m_b+m_s}\Bigg\{
\nonumber
\\
&& -\left[ (m_B + \mkone)(\varepsilon^{(\lambda)*}.q)
V_1^{\kone}(q^2) - (m_B -m_{k_1}) (\varepsilon_{(\lambda)}^* \cdot
p_B)
 V_2^{\kone}(q^2)
\right] \nonumber\\&& +2 \mkone (\varepsilon_{(\lambda)}^* \cdot
p_B) \left[
 V_3^{\kone}(q^2) - V_0^{\kone}(q^2)
\right]\Bigg\}, \label{formfactor3}\eea

The formfactors of $B\rightarrow K_1(1270)$ and $B\rightarrow
K_1(1400)$ can be expressed in terms of $B\rightarrow K_A$ and
$B\rightarrow K_B$ as follows(see \cite{Hatanaka:2008gu}):
\begin{eqnarray}
\pmatrix{
 \bra{\barkonel}|\sbar \gamma_\mu(1-\gamma_5) b |\ket{\barB} \cr
 \bra{\barkoneh}|\sbar \gamma_\mu(1-\gamma_5) b |\ket{\barB}}
&=& M \pmatrix{
 \bra{\barK_{1A}}|\sbar\gamma_\mu(1-\gamma_5) b|\ket{\barB} \cr
 \bra{\barK_{1B}}|\sbar\gamma_\mu(1-\gamma_5) b|\ket{\barB} },
\\
\pmatrix{ \bra{\barkonel}|\sbar \sigma_{\mu\nu}q^\nu(1+\gamma_5) b
|\ket{\barB} \cr \bra{\barkoneh}|\sbar
\sigma_{\mu\nu}q^\nu(1+\gamma_5) b |\ket{\barB} } &=& M \pmatrix{
 \bra{\barK_{1A}}|\sbar\sigma_{\mu\nu}q^\nu(1+\gamma_5) b|\ket{\barB} \cr
 \bra{\barK_{1B}}|\sbar\gamma_{\mu\nu}q^\nu(1+\gamma_5) b|\ket{\barB} }
,
\end{eqnarray}
using  the mixing matrix $M$ being given in Eq.~\eqref{mixing} the
formfactors $A^\kone,V_{0,1,2}^\kone$ and $T_{1,2,3}^\kone$ can be
written as follows:
\begin{eqnarray}
\pmatrix{
 A^{\konel}/(m_B + m_{\konel}) \cr
 A^{\koneh}/(m_B + m_{\koneh})}
&=& M \pmatrix{
 A^{\konea}/(m_B + m_{\konea}) \cr
 A^{\koneb}/(m_B + m_{\koneb})},
\\
\pmatrix{ (m_B+m_{\konel}) V_1^{K_1(1270)} \cr (m_B+m_{\koneh})
V_1^{K_1(1400)}} &=& M \pmatrix{ (m_B+m_{\konea})V_1^{K_{1A}} \cr
(m_B+m_{\koneb})V_1^{K_{1B}}},
\\
\pmatrix{ V_2^{K_1(1270)}/(m_B + m_{\konel}) \cr
V_2^{K_1(1400)}/(m_B + m_{\koneh})} &=& M \pmatrix{
V_2^{K_{1A}}/(m_B + m_{\konea}) \cr V_2^{K_{1B}}/(m_B +
m_{\koneb})},
\\
\pmatrix{ m_{\konel} V_0^{K_1(1270)} \cr m_{\koneh} V_0^{K_1(1400)}}
&=& M \pmatrix{ m_{\konea} V_0^{K_{1A}} \cr m_{\koneb}
V_0^{K_{1B}}},
\\
\pmatrix{ T_1^{K_1(1270)} \cr T_1^{K_1(1400)}}&=& M \pmatrix{
T_1^{K_{1A}} \cr T_1^{K_{1B}} },
\\
\pmatrix{ (m_B^2 - m_{\konel}^2) T_2^{K_1(1270)} \cr (m_B^2 -
m_{\koneh}^2) T_2^{K_1(1400)}}&=& M \pmatrix{ (m_B^2 - m_{\konea}^2)
T_2^{K_{1A}} \cr (m_B^2 - m_{\koneb}^2) T_2^{K_{1B}} },
\\
\pmatrix{ T_3^{K_1(1270)} \cr T_3^{K_1(1400)}} &=& M \pmatrix{
T_3^{K_{1A}} \cr T_3^{K_{1B}} },
\end{eqnarray}
where it is supposed that $p^\mu_{\konel,\koneh} \simeq
p^\mu_{\konea} \simeq p^\mu_{\koneb}$\cite{Hatanaka:2008gu}. These
formfactors within light-cone QCD sum rule (LCQSR) are estimated in
\cite{Yang:2008xw}.

Thus the matrix element for $B\to\kone\lpm$ in terms of formfacto is
given by
\begin{eqnarray}\label{ampl}
{\cal M} &=& \frac{G_F \alpha_{\rm em}}{2\sqrt{2}\pi} V_{ts}^*
V_{tb}^{}\, m_B \cdot (-i)\nnb \\ &&\Bigg\{
  \T_\mu^{(\kone),1} \lbar \gamma^\mu \l
 +\T_\mu^{(\kone),2} \lbar \gamma^\mu \gamma_5 \l
 +\T^{(\kone),3} \lbar \l+\T^{(\kone),4}\lbar  \gamma_5 \l\Bigg\},
\end{eqnarray}
where
\begin{eqnarray}
\T_\mu^{(\kone),1} &=&
 \A^\kone(\hats) \epsilon_{\mu\nu\rho\sigma}
 \varepsilon^{*\nu} \hatp_B^\rho \hatp_\kone^\sigma
-i \B^\kone(\hats)\varepsilon^{*}_\mu \nonumber\\&& +i
\C^\kone(\hats)( \varepsilon^{*} \cdot \hatp_B) \hatp_\mu +i
\D^\kone(\hats)( \varepsilon^{*} \cdot \hatp_B) \hatq_\mu,
\\
\T_\mu^{(\kone),2} &=&
 \E^\kone(\hats) \epsilon_{\mu\nu\rho\sigma}
 \varepsilon^{*\nu} \hatp_B^\rho \hatp_\kone^\sigma
-i \F^\kone(\hats)\varepsilon^{*}_\mu \nonumber\\&& +i
\G^\kone(\hats)( \varepsilon^{*} \cdot \hatp_B) \hatp_\mu +i
\H^\kone(\hats)( \varepsilon^{*} \cdot \hatp_B) \hatq_\mu,
\\
\T^{(\kone),3} &=&i {\cal I}_1^\kone(\hats)
\frac{(\varepsilon^{(\lambda)*}.\hat{q})}{1+\hat{m_s}} +i {\cal
J}_1^\kone(\hats)
\frac{(\varepsilon^{(\lambda)*}.\hatp_B)}{1+\hat{m_s}}
\\
\T^{(\kone),4} &=&i {\cal I}_2^\kone(\hats)
\frac{(\varepsilon^{(\lambda)*}.\hat{q})}{1+\hat{m_s}} +i {\cal
J}_2^\kone(\hats)
\frac{(\varepsilon^{(\lambda)*}.\hatp_B)}{1+\hat{m_s}}
\end{eqnarray}
with
 $\hatp = p/m_B$,
 $\hatp_B = p_B/m_B$,
 $\hatq= q/m_B$,
  $\hat{m}_s= m_s/m_B$,and
 $p = p_B + p_\kone$,
 $q = p_B - p_\kone = p_{\ell^+} + p_{\ell^-} $.
Here $\A^\kone(\hats), \cdots, \H^\kone(\hats)$ are defined by
\begin{eqnarray}
\A^\kone(\hats) &=& \frac{2}{1+\sqrt{\hat{r}_{K_1}}} c_9^{\eff}
(\hats) A^\kone(\hats) + \frac{4\hatm_b}{\hats} c_7^\eff
T^\kone_1(\hats), \label{Eq:A}
\\
\B^\kone(\hats) &=& (1+\sqrt{\hat{r}_{K_1}})\left[
 c_9^\eff (\hats) V_1^\kone(\hats)
 + \frac{2\hatm_b}{\hats} (1-\sqrt{\hat{r}_{K_1}})c_7^\eff T^\kone_2(\hats)
\right],
\\
\C^\kone(\hats) &=& \frac{1}{1-\hat{r}_{K_1}} \left[
 (1-\sqrt{\hat{r}_{K_1}}) c_9^\eff(\hats) V_2^\kone (\hats) + 2\hatm_b c_7^\eff
 \left(
  T_3^\kone(\hats) + \frac{1-\sqrt{\hat{r}_{K_1}}^2}{\hats} T_2^\kone(\hats)
 \right)
\right],
\nonumber\\
\\
\D^\kone(\hats) &=& \frac{1}{\hats} \biggl[
 c_9^\eff(\hats) \left\{(1+\sqrt{\hat{r}_{K_1}}) V_1^\kone(\hats)
  - (1-\sqrt{\hat{r}_{K_1}}) V_2^\kone(\hats)
  - 2\sqrt{\hat{r}_{K_1}} V_0^\kone(\hats) \right\}
\nonumber\\&&
  - 2\hatm_b c_7^\eff T_3^\kone(\hats)
\biggr],
\\
\E^\kone(\hats) &=& \frac{2}{1+\sqrt{\hat{r}_{K_1}}} c_{10}
A^\kone(\hats),
\\
\F^\kone(\hats) &=& (1 + \sqrt{\hat{r}_{K_1}}) c_{10}
V_1^\kone(\hats),
\\
\G^\kone(\hats) &=& \frac{1}{1 + \sqrt{\hat{r}_{K_1}}} c_{10}
V_2^\kone(\hats),
\\
\H^\kone(\hats) &=& \frac{1}{\hats} c_{10} \left[
 (1+\sqrt{\hat{r}_{K_1}}) V_1^\kone(\hats)
 - (1-\sqrt{\hat{r}_{K_1}}) V_2^\kone(\hats) - 2\sqrt{\hat{r}_{K_1}} V_0^\kone(\hats)
\right], \label{Eq:H}
\\
{\cal I}_1^\kone(\hats)&=& -C_{Q_1} (1 + \sqrt{\hat{r}_{K_1}})
V_1^{\kone}(\hats) \\
{\cal J}_1^\kone(\hats)&=&C_{Q_1} \{(1 +\sqrt{\hat{r}_{K_1}})
 V_2^{\kone}(\hats)+2
 \sqrt{\hat{r}_{K_1}}[V_3^{\kone}(\hats)-V_0^{\kone}(\hats)]\}\\
{\cal I}_2^\kone(\hats)&=&{\cal I}_1^\kone(\hats)(C_{Q_2}\rightarrow
C_{Q_1}),\,\,\,\,\,{\cal J}_2^\kone(\hats)={\cal
J}_1^\kone(\hats)(C_{Q_2}\rightarrow C_{Q_1})
\end{eqnarray}
with $\hat{r}_{K_1} = m^2_\kone/m^2_B$ and  $\hats = q^2/m_B^2$.

The dilepton invariant mass spectrum of the lepton pair for the
$\barB\to\barkone\lpm$ decay is given by
\begin{eqnarray}
\frac{d \Gamma(\barB\to\barkone\lpm)}{d \hats} = \frac{G_F^2
\alphaem^2 m_B^5}{2^{12}\pi^5}
 \left|V_{tb}V_{ts}^*\right|^2 v\sqrt{\lambda}\Delta(\hats)
\end{eqnarray}
where $v=\sqrt{1-4\hat{m}_\ell^2/\hats}$, $\lambda = 1 +
\hat{r}_{\kone}^2 + \hats^2 - 2\hats - 2\hat{r}_{\kone} (1+\hats)$
and
 \bea\label{dgds1}\nnb \Delta(\hats)&=&\frac{8Re[{\cal F}{\cal H}^*]
\hat{m}_{\ell}^2 \lambda}{\hat{r}_{K_1}}+\frac{8Re[{\cal G}{\cal
H}^*]
\hat{m}_{\ell}^2(-1+\hat{r}_{K_1})\lambda}{\hat{r}_{K_1}}-\frac{8|{\cal
H}|^2 \hat{m}_{\ell}^2 \hat{s} \lambda}{\hat{r}_{K_1}}
\\ \nnb &-&
\frac{2Re[{\cal B}{\cal
C}^*](-1+\hat{r}_{K_1}+\hat{s})(3+3\hat{r}_{K_1}^2-6\hat{s}+3\hat{s}^2-6\hat{r}_{K_1}(1+\hat{s})-v^2\lambda)}{3\hat{r}_{K_1}}
\\ \nnb &-&
\frac{|{\cal C}|^2
\lambda(3+3\hat{r}_{K_1}^2-6\hat{s}+3\hat{s}^2-6\hat{r}_{K_1}(1+\hat{s})-v^2\lambda)}{3\hat{r}_{K_1}}
\\ \nnb &-&
\frac{|{\cal G}|^2
\lambda(3+3\hat{r}_{K_1}^2+12\hat{m}_{\ell}^2(2+2\hat{r}_{K_1}-\hat{s})-6\hat{s}+3\hat{s}^2-6\hat{r}_{K_1}(1+\hat{s})-v^2\lambda)}{3\hat{r}_{K_1}}
\\ \nnb &+&
\frac{|{\cal
F}|^2(-3-3\hat{r}_{K_1}^2+6\hat{r}_{K_1}(1+16\hat{m}_{\ell}^2-3\hat{s})+6\hat{s}-3\hat{s}^2+v^2\lambda)}{3\hat{r}_{K_1}}
\\ \nnb &+&
\frac{|{\cal
B}|^2(-3-3\hat{r}_{K_1}^2+6\hat{s}-3\hat{s}^2-6\hat{r}_{K_1}(-1+8\hat{m}_{\ell}^2+3\hat{s})+v^2\lambda)}{3\hat{r}_{K_1}}
\\ \nnb &+&
\frac{2}{3\hat{r}_{K_1}}Re[{\cal F}{\cal
G}^*](12\hat{m}_{\ell}^2\lambda-(-1+\hat{r}_{K_1}+\hat{s})
(3+3\hat{r}_{K_1}^2-6\hat{s}+3\hat{s}^2-6\hat{r}_{K_1}(1+\hat{s})-v^2\lambda))
\\ \nnb &+&
|{\cal
A}|^2(-4\hat{m}_{\ell}^2\lambda-\frac{\hat{s}}{3}(3+3\hat{r}_{K_1}^2-6\hat{s}+3\hat{s}^2-6\hat{r}_{K_1}(1+\hat{s})+v^2\lambda))
\\ \nnb &+&
|{\cal
E}|^2(4\hat{m}_{\ell}^2\lambda-\frac{\hat{s}}{3}(3+3\hat{r}_{K_1}^2-6\hat{s}+3\hat{s}^2-6\hat{r}_{K_1}(1+\hat{s})+v^2\lambda))\nnb
\\ &+&\lambda\{ \frac{\left(4 \hat{m}_\ell^2-\hat{s}\right)   |{\cal I}_1|^2}{\hat{r}_{K_1}}
+\frac{|{\cal J}_1|^2 \left(4 \hat{m}_\ell^2-\hat{s}\right) }
{\hat{r}_{K_1}}+\frac{2Re[{\cal I}_1{\cal J}_1^\ast] \left(4
\hat{m}_\ell^2-\hat{s}\right)}{\hat{r}_{K_1}}
   -\frac{|{\cal I}_2|^2 \hat{s}  }{\hat{r}_{K_1}}\nnb\\ &-&\frac{|{\cal J}_2|^2 \hat{s}}{\hat{r}_{K_1}}
   -\frac{2Re[{\cal I}_1{\cal J}_1^\ast] \hat{s}
   }{\hat{r}_{K_1}}+\frac{4
   Re[{\cal H I}_2^\ast]\hat{m}_\ell \hat{s}  }{\hat{r}_{K_1}}+\frac{4 Re[{\cal HJ}_2^\ast] \hat{m}_\ell \hat{s}
   }{\hat{r}_{K_1}}-\frac{4
   Re[{\cal F I}_2^\ast] \hat{m}_\ell  }{\hat{r}_{K_1}}\nnb \\ &-&\frac{4 Re[{\cal F J}_2^\ast] \hat{m}_\ell  }{\hat{r}_{K_1}}
   -\frac{4 Re[{\cal GI}_2^\ast] \hat{m}_\ell
   (\hat{r}_{K_1}-1)  }{\hat{r}_{K_1}}-\frac{4 Re[{\cal GJ}_2^\ast] \hat{m}_\ell (\hat{r}_{K_1}-1)
   }{\hat{r}_{K_1}}\}
 \eea

The normalized differential forward-backward asymmetry of the
$\barB\to\barkone\lpm$ decay is defined by

\bea {\cal
A}_{FB}(\hats)&=&\frac{\int_{0}^1\Gamma(\hats,\cos(\theta))d
\cos(\theta)-\int_{-1}^0\Gamma(\hats,\cos(\theta))d
\cos(\theta)}{\int_{0}^1\Gamma(\hats,\cos(\theta))d
\cos(\theta)+\int_{-1}^0\Gamma(\hats,\cos(\theta))d \cos(\theta)}
\eea Using the definition mentioned above we calculate the
normalized differential forward-backward asymmetry(FBA). The result
is as follows:
\begin{eqnarray}
 {\cal A}_{FB}(\hats)&=&\frac{v \sqrt{\lambda} }{\hat{r}_{K_1}\Delta}\Bigg\{ 2(Re[{\cal AF^\ast}]+Re[{\cal BE^\ast}])\hat{r}_{K_1}\hat{s}
 +\hat{m}_\ell Re[{\cal B}({\cal I}_1+{\cal J}_1)^\ast](-1+\hat{r}_{K_1}+\hat{s})\nnb \\
 &+&\hat{m}_\ell Re[{\cal C}({\cal I}_1+{\cal J}_1)^\ast]\lambda\Bigg\}
\end{eqnarray}
Note that  the  pseudoscalar structure existing in the decay
amplitude(Eq.~\ref{ampl}) can affect the branching ratio, the same
structure don't contribute to the expression of the FBA. Thus, the
study of FBA is complimentary to the study of branching ratio in
order to extract the information about the nature of interactions in
SUSY models.
\section{Numerical results}

In this section, we present the branching ratio and FB asymmetry for
the $B \rar K_1(1270) \ell^+ \ell^-$ decay for muon and tau leptons.
The main input parameters are the form factors for which we use the
results of light cone  QCD sum rules(LCQCD)~\cite{Yang:2008xw}. We
use the parameters given in Tables~\ref{input} and \ref{tab:FFinLF}
in our numerical analysis.
\begin{table}[tbp]
\caption{Input parameters}\label{input}
        \begin{center}
        \begin{tabular}{|l|l|}
        \hline
        \multicolumn{1}{|c|}{Parameter} & \multicolumn{1}{|c|}{Value}     \\
        \hline \hline
         $\alpha_{s}(m_Z)$                   & $0.119$  \\
        $\alpha_{em}$                   & $1/129$\\
        $m_{\konel}$                   & $1.270$ (GeV)\cite{Yao:2006px} \\
        $m_{\koneh}$                   & $1.403$ (GeV) \cite{Yao:2006px}\\
        $m_{\konea} $                  & $1.31$ (GeV) \cite{Yang:2007zt}\\
        $m_{\koneb} $                  & $1.34$ (GeV) \cite{Yang:2007zt}\\
        $m_{b}$                   & $4.8$ (GeV) \\
        $m_{\mu}$                   & $0.106$ (GeV) \\
        $m_{\tau}$                  & $1.780$ (GeV) \\
        \hline
        \end{tabular}
        \end{center}
\label{input}
\end{table}
The values of the form factors at $q^2=0$ are given in table
3\cite{Yang:2008xw}
\begin{table}[t]
\caption{Formfactors for $B\to K_{1A},K_{1B}$ transitions obtained
in the LCQSR calculation \cite{Yang:2008xw} are fitted to the
3-parameter form in Eq. (\ref{e8425}).} \label{tab:FFinLF}
\begin{tabular}{clll|clll}
      ~~~~$F$~~~~~~
    & ~~~~~$F(0)$~~~~~
    & ~~~$a$~~~
    & ~~~$b$~~
    & ~~~~$F$~~~~~~
    & ~~~~~$F(0)$~~~~~
    & ~~~$a$~~~
    & ~~~$b$~~
 \\
    \hline
$V_1^{BK_{1A}}$
    & $0.34\pm0.07$
    & $0.635$
    & $0.211$
&$V_1^{BK_{1B}}$
    & $-0.29^{+0.08}_{-0.05}$
    & $0.729$
    & $0.074$
    \\
$V_2^{BK_{1A}}$
    & $0.41\pm 0.08$
    & $1.51$
    & $1.18~~$
&$V_2^{BK_{1B}}$
    & $-0.17^{+0.05}_{-0.03}$
    & $0.919$
    & $0.855$
    \\
$V_0^{BK_{1A}}$
    & $0.22\pm0.04$
    & $2.40$
    & $1.78~~$
&$V_0^{BK_{1B}}$
    & $-0.45^{+0.12}_{-0.08}$
    & $1.34$
    & $0.690$
    \\
$A^{BK_{1A}}$
    & $0.45\pm0.09$
    & $1.60$
    & $0.974$
&$A^{BK_{1B}}$
    & $-0.37^{+0.10}_{-0.06}$
    & $1.72$
    & $0.912$
    \\
$T_1^{BK_{1A}}$
    & $0.31^{+0.09}_{-0.05}$
    & $2.01$
    & $1.50$
&$T_1^{BK_{1B}}$
    & $-0.25^{+0.06}_{-0.07}$
    & $1.59$
    & $0.790$
    \\
$T_2^{BK_{1A}}$
    & $0.31^{+0.09}_{-0.05}$
    & $0.629$
    & $0.387$
&$T_2^{BK_{1B}}$
    & $-0.25^{+0.06}_{-0.07}$
    & $0.378$
    & $-0.755$
    \\
$T_3^{BK_{1A}}$
    & $0.28^{+0.08}_{-0.05}$
    & $1.36$
    & $0.720$
&$T_3^{BK_{1B}}$
    & $-0.11\pm 0.02$
    & $-1.61$
    & $10.2$
\end{tabular}

\end{table}

The best fit for the $q^2$ dependence of the form factors can be
written in the following form: \bea \label{e8425} f_i(\hat{s}) =
{f_i(0)\over 1 - a_i \hat{s} + b_i \hat{s}^2}~, \eea  The values of
the parameters $f_i(0)$, $a_i$ and $b_i$   are given in Table 3.

The mixing angle  $\thetaK$ was estimated to be $|\thetaK| \approx
34\degree\vee 57\degree$ in Ref. \cite{Suzuki:1993yc}, $35\degree
\leq |\thetaK| \leq 55\degree$ in Ref.~\cite{Burakovsky:1997ci},
$|\thetaK|= 37\degree \vee 58\degree$ in Ref.~\cite{Cheng:2003bn},
and $\thetaK= -(34 \pm 13)\degree $ in \cite{Hatanaka:2008gu,
Hatanaka:2008xj}. In this study, we  use the results of
Ref.\cite{Hatanaka:2008gu,Hatanaka:2008xj} for numerical
calculations, where we take $\thetaK=-34^\circ$.

The new Wilson coefficients $C_{Q_1}$ and $C_{Q_2}$ describes in
terms of   masses of sparticles i.e., chargino-up-type squark and
NHBs,  $\tan(\beta)$ which is defined as the ratio of the two vacuum
values of the 2 neutral Higgses and $\mu$  which has the dimension
of a mass, corresponding to a mass term mixing the 2 Higgses
doublets. Note that $\mu$ can be positive or negative. Depending on
the magnitude and sign of these parameters one can consider many
options in the parameter space, but experimental results i.e., the
rate of $b\rightarrow s \gamma$ and $b\rightarrow s \ell^+ \ell^-$
constrain us to consider the following options
\begin{itemize}
\item {SUSY I: $\mu$ takes negative value, $C_7$ changes its sign and
contribution of NHBs are neglected.}
\item {SUSY II: $\tan(\beta)$ takes large values while the mass of
superpartners are small i.e., few hundred GeV.}
\item {SUSY III: $\tan(\beta)$ is large and the masses of superpartners are relatively
large, i.e., about 450 GeV or more.}
\end{itemize}
The numerical values of Wilson coefficients  used in our analysis
are borrowed from Ref. \cite{Aslam:2008hp,aslam12} and collected in
Tables 4, and 5.

The numerical results for the decay rates and  FBAs are presented in
Figs. 1-4. Fig.~1 describes the differential decay rate of $B \rar
K_1(1270) \mu^+ \mu^-$, from which one can see that the
supersymmetric effects are quite significant(about twice of SM ) for
SUSY I and SUSY II models in the low momentum transfer regions,
whereas these effects are small for SUSY III case. The reason for
the increase of differential decay width in SUSY I model is the
relative change in the sign of $C^{eff}_7$ which gives dominant
contribution in the low momentum transfer regions(look at the factor
of $1/q^2$ in the Eq. \ref{e1}), while the large change in SUSY II
model is owing to the contribution of the NHBs. The same effects can
also be seen for the $\tau$ channel( see fig. 2). Fig. 3 describes
the FBA of $B \rar K_1(1270) \mu^+ \mu^-$, from which one can see
that except SUSY III the supersymmetric effects are drastic in the
low momentum transfer regions. In SUSY I and SUSY II models, the
sign of $C^{eff}_7$ and $C^{eff}_9$ become the same, hence, the zero
point of the FBAs disappears. Though, in the SUSY III model FBA
passes from the zero but this zero position shifts to the right from
that of the SM value due to the contribution from the NHBs. FBA is
suppressed with the supersymmetric effects. The suppression is much
more in the SUSY II model than the others(see fig. 4).
\begin{table}[tbh]
\begin{center}
\begin{tabular}{ccccccc}
\hline Wilson Coefficients & $C_{7}^{eff}$ &  $C_{9}$ &  $C_{10}$ \\
\hline SM & $-0.313$ &  $4.334$ &  $-4.669$\\
\hline SUSY I & $+0.3756$ &  $4.7674$ & $-3.7354$  \\
\hline SUSY II & $+0.3756$ & $4.7674$ &  $-3.7354$
\\ \hline\hline
SUSY III & $-0.3756$ & $4.7674$ & $-3.7354$ \\
\hline\hline
\end{tabular}%
\end{center}
\caption{{}Wilson Coefficients in SM and different SUSY models
without NHBs contributions.}%
\end{table}

\begin{table}[tbh]
\label{NHB}%
\begin{center}
\begin{tabular}{ccccc}
\hline Wilson Coefficients & $C_{Q_{1}}$ & $C_{Q_{2}}$   \\ \hline
SM & $0$  & $0$ \\ \hline SUSY I & $0$ &  0  \\
\hline SUSY II & $6.5\left( 16.5\right) $ & $-6.5\left( -16.5\right)
$
\\ \hline\hline
SUSY III & $1.2\left( 4.5\right) $ &  $-1.2\left( -4.5\right) $  \\
\hline\hline
\end{tabular}%
\end{center}
\caption{{}Wilson coefficient corresponding to NHBs contributions
within  SUSY I, II and III models~\cite{Aslam:2008hp}. The values in the bracket are for the $\protect%
\tau $.}
\end{table}

To sum up, we study the semileptonic rare $B \rar K_1^\ast(1270)
\ell^+ \ell^-$ decay in the supersymmetric theories. We show that
the branching ratio and FBA are very sensitive to the SUSY
parameters. The branching ratio is enhanced up to one order of
magnitude with respect to the corresponding SM values. The magnitude
and sign of FBA show quite a significant discrepancy with respect to
the SM values. The results of this study can be used to indirect
search for the SUSY effects in future planned experiments at LHC.

\section*{Acknowledgments}
The authors thank T. M. Aliev for his useful discussions.

\newpage

\begin{figure}
\vskip 1.5 cm
    \includegraphics{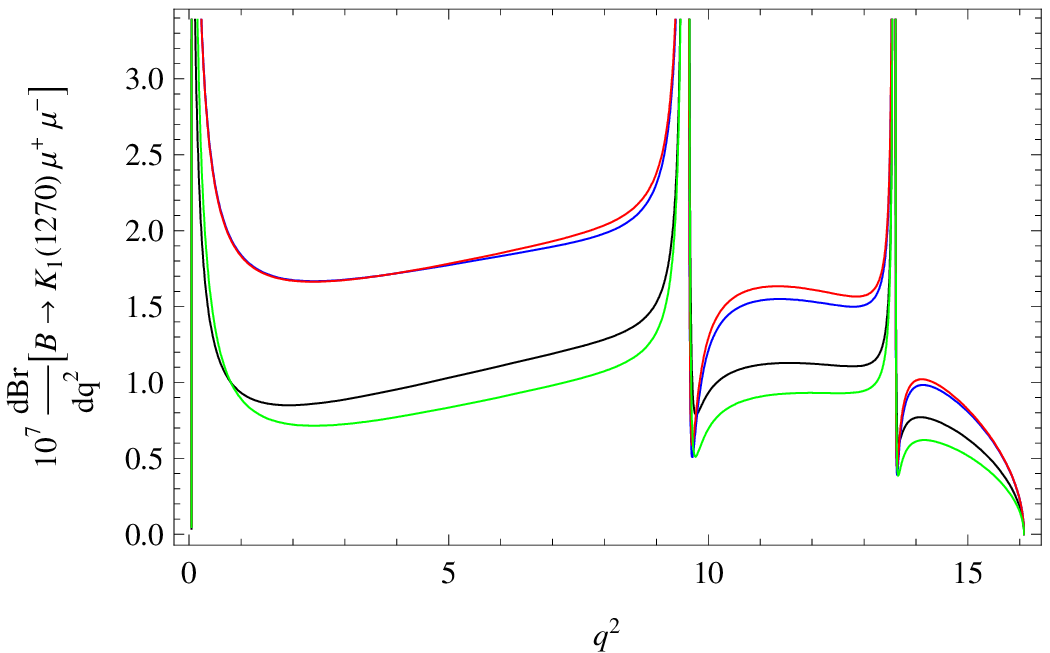}
\vskip 6 cm \caption{Branching ratio of $B \rar K_1^\ast(1270) \mu^+
\mu^-$ decay. The black, blue, red and green lines correspond to SM,
SUSY I, SUSY II, SUSY III models, respectively.  }
\end{figure}

\begin{figure}
\vskip 1.5 cm
    \includegraphics{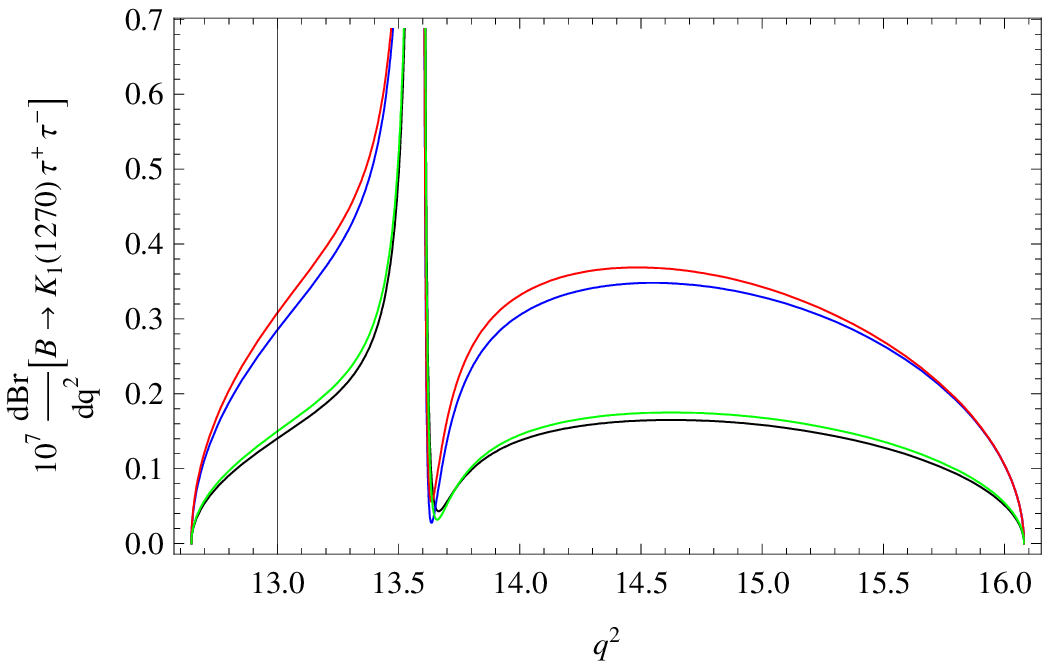}
\vskip 6 cm \caption{The same as Fig.~1 but for $\tau$ channel}
\end{figure}

\begin{figure}
\vskip 1.5 cm
    \includegraphics{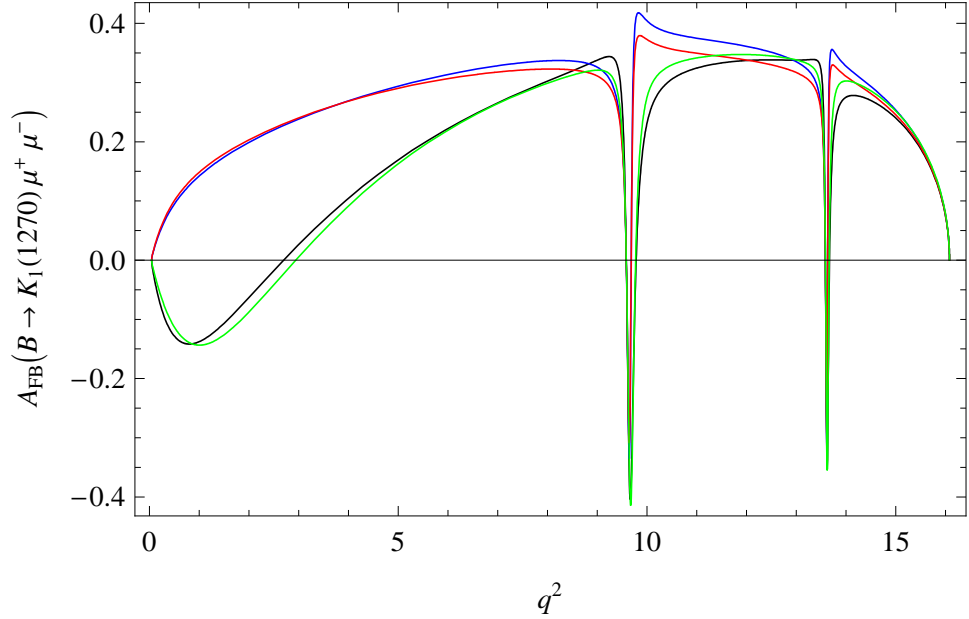}
\vskip 6 cm \caption{FBA of $B \rar K_1^\ast(1270) \mu^+ \mu^-$
decay. The black, blue, red and green lines correspond to SM, SUSY
I, SUSY II, SUSY III models, respectively.}
\end{figure}

\begin{figure}
\vskip 1.5 cm
    \includegraphics{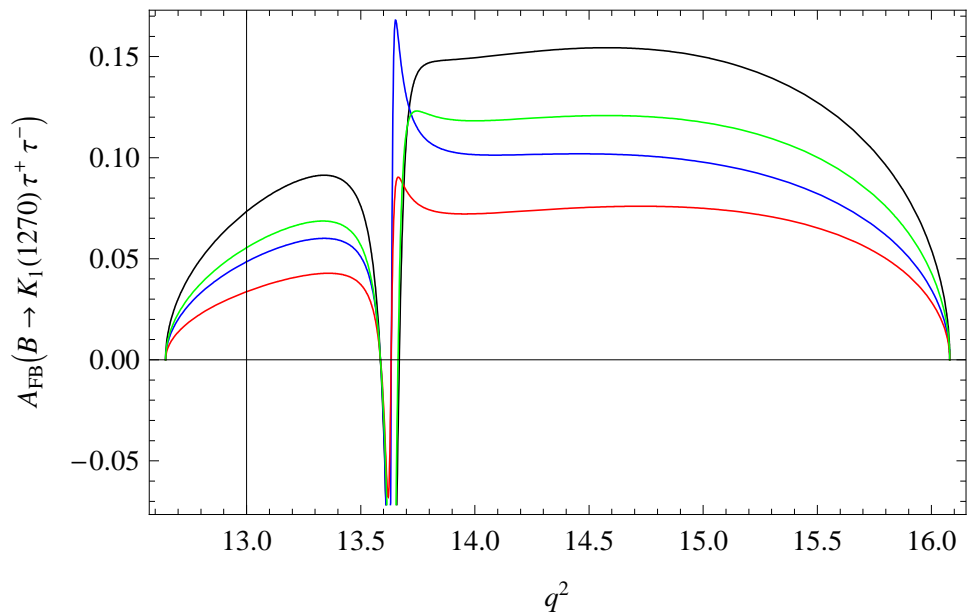}
\vskip 6 cm \caption{The same as Fig.~3 but for $\tau$ channel}
\end{figure}

\end{document}